\documentclass[12pt]{article}

\usepackage{times}
\usepackage{graphicx}
\usepackage{overcite}
\makeatletter
\renewcommand{\@biblabel}[1]{{#1}.}
\makeatother

\def\ä{\"{a}}
\def\ü{\"{u}}
\def\ö{\"{o}}
\def\Ä{\"{A}}
\def\Ü{\"{U}}
\def\Ö{\"{O}}

\begin{document}

\begin{center}
 {{\LARGE\bf Atomic force microscopy's path to atomic resolution\\}}

\vspace*{1.0cm}

{\bf Franz J. Giessibl}

\end{center}

\vspace*{1.0cm}

\date{\today}

\baselineskip24pt

{\large We review the progress in the spatial resolution of atomic
force microscopy (AFM) in vacuum. After an introduction of the
basic principle and a conceptual comparison to scanning tunneling
microscopy, the main challenges of AFM and the solutions that have
evolved in the first twenty years of its existence are outlined.
Some crucial steps along the AFM's path towards higher resolution
are discussed, followed by an outlook on current and future
applications.

\emph{Review Feature, submitted to Materials Today Feb 10 2005,
2nd revised version}}

\vspace*{1cm}

\baselineskip24pt

{ {\small Experimentalphysik VI, EKM,

Institute of Physics, Augsburg University,

86135 Augsburg, Germany,

Email: franz.giessibl@physik.uni-augsburg.de}
} \baselineskip12pt

\baselineskip24pt

{\bf Atomic force microscopy, invented\cite{Binnig:1986a} and also
introduced\cite{Binnig:1986b} in 1985/86, can be viewed as a
mechanical profiling technique that generates three-dimensional
maps of surfaces by scanning a sharp probe attached to a
cantilever over a surface. The forces that act between the tip of
the cantilever and the sample are used to control the vertical
distance. AFM's potential to reach atomic resolution was foreseen
in the original scientific publication \cite{Binnig:1986b}, but
for a long time, the spatial resolution of AFM was inferior to the
resolution capability of its parent, scanning tunneling microscopy
(STM). The resolution limits of STM and AFM are given by the
structural properties of the atomic wavefunctions of the probe tip
and the sample. STM is sensitive to the most loosely bonded
electrons with an energy at the Fermi level while AFM responds to
all electrons, including core electrons. Because the electrons at
the Fermi level are spatially less confined than core electrons
that contribute to AFM images, in theory AFM should be able to
achieve even greater spatial resolution than STM. Today,
experimental evidence emerges where in simultaneous AFM/STM
studies, AFM images reveal even finer structural details than
simultaneously recorded STM images. The experimental advances that
made high-resolution AFM possible started with the introduction of
frequency modulation AFM (FM-AFM), where the cantilever oscillates
at a fixed amplitude and the frequency is used as a feedback
signal. Early implementations of FM-AFM utilized silicon
cantilevers with a typical spring constant of 10\,N/m that
oscillate with an amplitude on the order of 10 nanometers. The
spatial resolution could be increased by the introduction of
quartz cantilevers with a stiffness on the order of 1\,kN/m,
allowing the use of sub-nm amplitudes. The direct evaluation of
higher harmonics in the cantilever motion has enabled a further
increase in spatial resolution. Because AFM can image insulators
as well as conductors, it is now a powerful complement to STM for
atomically resolved surface studies. Immediate applications of
high resolution AFM have been demonstrated in vacuum studies
relating to materials science, surface physics and -chemistry.
Some of the techniques developed for ultrahigh-vacuum AFM may be
applicable to increase AFM resolution in ambient or liquid
environments, such as required for studying biological or
technological specimens. }

\section{Introduction}

In terms of the operating principle, atomic force microscopy (AFM)
\cite{Binnig:1986a,Binnig:1986b} can be viewed as an extension of
the toddlers way of \lq grasping\rq{} the world by touching and
feeling as indicated in Figure 1 of Binnig and Rohrer's article
\lq{\it In touch with atoms}\rq \cite{Binnig:1999}, where a finger
profiles an atomic surface. Likewise, one could argue that stylus
profilometry is a predecessor of AFM. However, AFM and stylus
profilometry have as much in common as a candle and a laser. Both
of the latter generate light, and even though candles are
masterpieces of engineering\cite{Faraday:1861}, the laser is a
much more advanced technological device requiring a detailed
knowledge of modern quantum mechanics\cite{Siegman:1986}. While
stylus profilometry is an extension of human capabilities that
have been known for ages and works by classical mechanics, AFM
requires a detailed understanding of the physics of chemical
bonding forces and the technological prowess to measure forces
that are several orders of magnitude smaller than the forces
acting in profilometry. Only the spectacular spatial resolution of
scanning tunneling microscopy (STM) could trigger the hope that
the force acting between any STM tip and sample might lead to
atomic force microscopy capable of true atomic resolution. The
STM, established in 1981, is the first instrument that has allowed
to image surfaces with atomic resolution in real space
\cite{Binnig:1982,Hofer:2002}. The atomic imaging of the
7$\times$7 reconstruction of Si (111) by STM in
1983\cite{Binnig:1983} has later helped to solve one of the most
intriguing problems of surface science at that time and establish
the dimer-adatom-stacking fault model by Takayanagi et
al.\cite{Takayanagi:1985}. The capability of atomic resolution by
STM provided immediate evidence for the enormous value of this
instrument as a tool for surface scientists. STM can only be used
on conductive surfaces. Given that many surfaces of technological
interest are conducting or at least semiconducting, this may not
seem to be a severe shortcoming. One might think that an STM
should be capable of mapping the surface of a metallic surface at
ambient conditions. However, this is not feasible, because the
pervasive layer of oxides and other contaminants occurring at
ambient conditions prevents stable tunneling conditions.
Electrical conductivity is a necessary, but not a sufficient
condition for a surface to be imaged by STM with atomic
resolution, because surfaces need to be extremely clean on an
atomic level. Except for a few extremely inert surfaces such as
graphite, atomic resolution is only possible in an ultra-high
vacuum with a pressure on the order of 10$^{-8}$\,Pa and special
surface preparation. The invention of the AFM by
Binnig\cite{Binnig:1986a} and its introduction by Binnig, Quate
and Gerber\cite{Binnig:1986b} opened the possibility of obtaining
true atomic resolution on conductors \emph{and} insulators.
Indeed, it took only a short time after the AFM's invention before
apparently atomic resolution on conductors\cite{Binnig:1987a} and
insulators\cite{Albrecht:1987,MeyerG:1990,MeyerE:1990a} was
obtained. While these early results reproduced the periodic
lattice spacings of the samples that were studied, single defects
or step edges were not observed. Also, the forces that acted
between tip and sample were often orders of magnitudes larger than
the forces that a tip with a single front atom was expected to be
able to sustain. Therefore, it was commonly assumed that
\emph{many} tip atoms interacted with the surface at the same time
in these early experiments. The difference between \emph{apparent}
and \emph{true} atomic resolution of a tip with many atomic
contacts can be illustrated by a macroscopic example: When
profiling an egg crate with a single egg, its trajectory would
represent the overall periodicity of the crate as well as a dented
hump or a hole. However, when profiling one egg crate with another
egg crate, again its periodicity would be retained, but holes or
dented humps would pass undetected. A similar effect can occur
when an AFM tip probes a surface. As long as single defects, steps
or other singularities are not observed, a clear proof for
\emph{true} atomic resolution is not established. Even though
atomic resolution was hardly ever achieved in initial AFM
experiments, the technique was readily accepted and found many
technological and scientific applications. The installed base of
atomic force microscopes rapidly outnumbered their STM
counterparts. A recent survey\cite{Riordon:2003} about the ten
most highly cited publications of {\it Physical Review Letters}
ranks the original AFM publication\cite{Binnig:1986b} as number
four (4251 citations as of Mar 11 2005 [ISI]) -- in good company
with other breakthroughs in theoretical and experimental physics
that have shaped our scientific life. Most of these citations
refer to AFM where the spatial resolution is \lq only\rq{} in the
nanometer-range, but the large number proves the vast applications
of AFM. In spite of the rapid growth of AFM applications, matching
and even exceeding the spatial resolution of STM, its parent, had
to wait for new experimental developments.

\section{Challenges of atomic resolution AFM}

\begin{figure}[h]
\includegraphics[width=16cm]{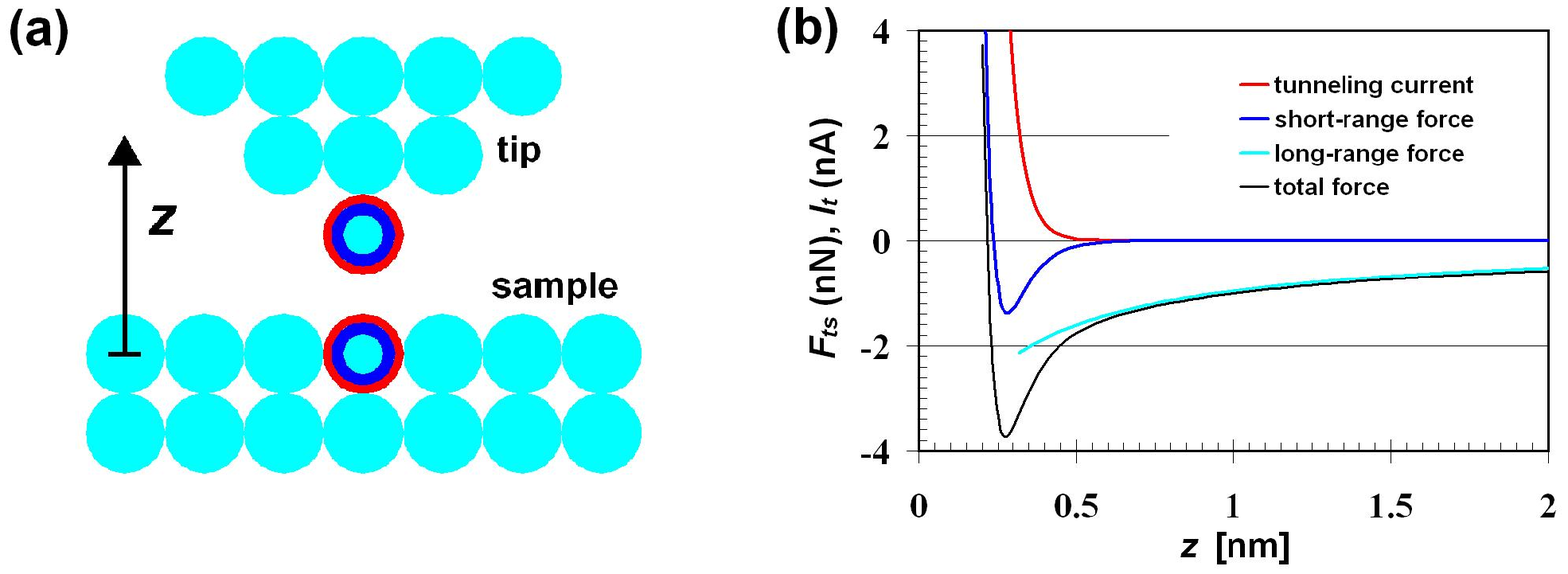}
\caption{(a) Schematic view of a tip and a sample in a scanning
tunneling microscope or atomic force microscope. The diameter of a
metal atom is typically 0.3\,nm. (b) Qualitative distance
dependence of tunneling current, long- and short-range forces. The
tunneling current increases monotonic with decreasing distance,
while the force reaches a minimum and increases for distances
below the bond length.\label{figtsIF}}
\end{figure}

The technological foundations for the feasibility of STM with
atomic resolution (theory of electron tunneling, mechanical
actuation with pico-meter precision, vacuum technology, surface-
and tip preparation, vibration isolation, ...) were probably
available a few decades before 1981, but it took the bold approach
by Gerd Binnig, Heinrich Rohrer, Christoph Gerber and Edi Weibel
to pursue atomic resolution in real space. Binnig and Rohrer were
rewarded with the Nobel Prize in Physics in 1986 (together with
Ernst Ruska, the inventor of Electron Microscopy) and the AFM
principle was published in the same year. The challenges of AFM
with true atomic resolution are even more daunting than the
hurdles that troubled STM. To start our discussion of the special
AFM challenges, we first look at the physics behind STM. Figure
\ref{figtsIF} (a) shows a schematic view of a sharp tip for STM or
AFM close to a crystalline sample and Fig. \ref{figtsIF} (b) is a
plot of the tunneling current and forces between tip and sample.
When tip and sample are conductive and a bias voltage is applied
between them, a tunneling current can flow. The red curve in Fig.
\ref{figtsIF} (b) shows the distance dependence of the tunneling
current $I_t$. The exponential decay of $I_t$ with distance at a
rate of approximately one order of magnitude per 100\,pm distance
increase is the key physical characteristic that makes atomic
resolution STM possible. Because of its strong decay rate, the
tunneling current is spatially confined to the front atom of the
tip and flows mainly to the sample atom next to it (indicated by
red circles in Fig. \ref{figtsIF} (a)). A second helpful property
of the tunneling current is its monotonic distance dependence. It
is easy to build a feedback mechanism that keeps the tip at a
constant distance: if the actual tunneling current is larger than
the setpoint, the feedback  needs to withdraw the tip and vice
versa. The tip sample force $F_{ts}$, in contrast, does not share
the helpful key characteristics of the tunneling current. First,
$F_{ts}$ is composed of long-range background forces depicted in
light-blue in Fig. \ref{figtsIF} (b) and originating from the
atoms colored light-blue in Fig. \ref{figtsIF} (a) and a
short-range component depicted in blue in Fig. \ref{figtsIF} (b)
and confined to the atoms printed in blue in Fig. \ref{figtsIF}
(a). Because the short-range force is not monotonic, it is
difficult to design a feedback loop that controls distance by
utilizing the force. A central task to perfect AFM is therefore
the isolation the front atom's force contribution and the creation
of a linear feedback signal from it.

Even if it was possible to isolate the short-range force, a more
basic problem needs to be solved first: how to measure small
forces. For example, commonly known force meters such as precise
scales are delicate and expensive instruments and even top models
rarely exceed a mass resolution of 100\,$\mu$g, corresponding to a
force resolution of 1\,$\mu$N. In addition, high-precision scales
take about one second to acquire a weight measurement so the
bandwidth is only 1\,Hz. The force meters in AFM, in contrast,
require a force resolution of at least a nano-Newton at a typical
bandwidth of 1\,kHz. Most force meters determine the deflection
$q'$ of a spring with given spring constant $k$ that is subject to
a force $F$ with $F=q'/k$. Measuring small spring deflections is
subject to thermal drift and other noise factors, resulting in a
finite deflection measurement accuracy $\delta q'$. The force
resolution is thus given by $\delta F=\delta q'/k$, and soft
cantilevers provide less noise in the force measurement. In
contact-mode AFM, where the tip feels small repulsive forces from
the sample surface, the cantilever should be softer than the bonds
between surface atoms (estimated at $\approx 10$\,N/m), otherwise
the sample deforms more than the cantilever \cite{Rugar:1990}.
Because of noise and stability considerations, spring constants
below 1\,N/m or so have been chosen for AFM in contact mode.
However, atomic forces are usually attractive in the distance
regime that is best suited for atomic resolution imaging
(approximately a few hundred pm before making contact), and soft
cantilevers suffer from a "jump-to-contact" phenomenon, i.e. when
approaching the surface, the cantilever snaps towards the surface
ended by an uncontrolled landing. While true atomic resolution by
contact-mode AFM has been demonstrated on samples that are
chemically inert \cite{Giessibl:1992b,Ohnesorge:1993}, this method
is not feasible for imaging reactive surfaces where
strong attractive short-range forces act. The long-range attractive forces
have been compensated in these experiments by pulling at the
cantilever (negative loading force) after
jump-to-contact\cite{Giessibl:1992b} or by immersing cantilever
and sample in water to reduce the van-der-Waals
attraction\cite{Ohnesorge:1993}. Howald et al. \cite{Howald:1995}
could partially solve the reactivity problem by passivating the
reactive Si tip with a thin layer of poly-tetra-fluor-ethylen
(teflon). The unit cell of Si(111)-(7$\times$7) was resolved, but
atomic resolution was not reported with that method of tip
passivation.

In summary, AFM shares the challenges that are already known from
STM and uses many of its design features (actuators, vibration
isolation etc.), but nature has posed four extra problems for
atomic resolution AFM: 1. Jump-to-contact, 2. Non-monotonic short
range forces, 3. Strong long-range background forces and 4.
Instrumental noise in force measurements.

\section{Frequency modulation atomic force microscopy}

Dynamic AFM modes \cite{Martin:1987,Albrecht:1991,Duerig:1992} help to
alleviate two of the four major AFM challenges. Jump-to-contact
can be prevented by oscillating the cantilever at a large enough
amplitude $A$ such that the withdrawing force on the cantilever
given by $k\times A$ is larger than the maximal attractive
force\cite{Giessibl:1997b}. Because the noise in cantilever
deflection measurements has a component that varies in intensity
inversely with frequency ($1/f$-noise), dynamic AFM modes are less
subject to noise than quasistatic operating modes. Non-monotonic
interactions and strong long-range contributions are still
present.

\begin{figure}
\includegraphics[width=16cm]{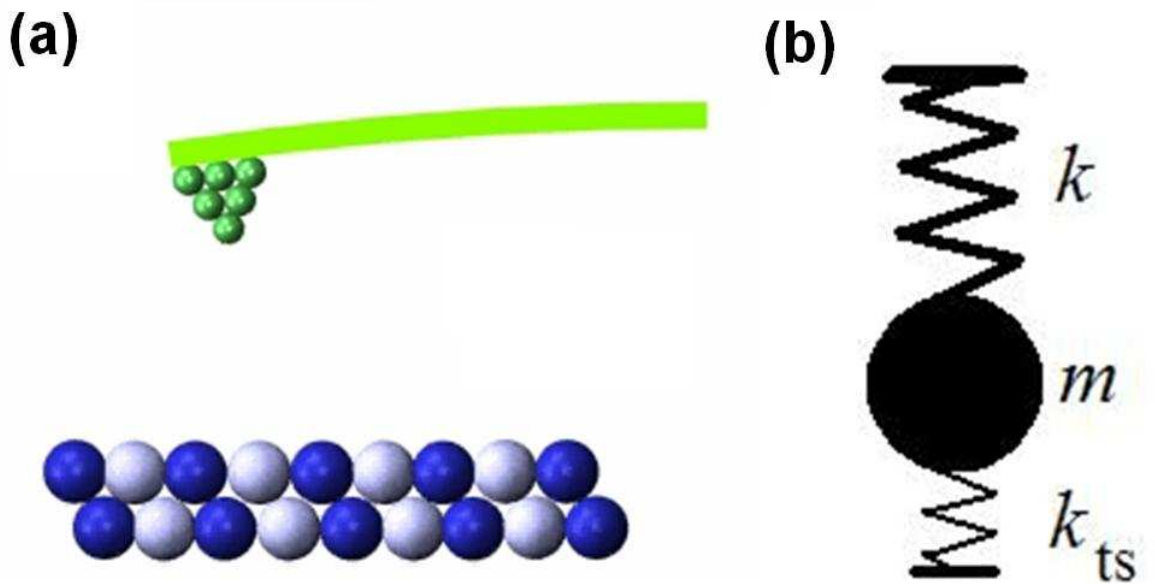}
\caption{(a) Schematic view of a vibrating tip close to a a sample
in a dynamic atomic force microscope. The forces that act between
the tip and the sample $F_{ts}$ cause a detectable change in the
oscillation properties of the cantilever. (b) Mechanical
equivalent of (a). The free cantilever with stiffness $k$ and
effective mass $m$ can be treated as a harmonic oscillator with an
eigenfrequency $f_0=(k/m)^{1/2}/(2\pi)$. The bond between tip and
sample with its stiffness $k_{ts}$ alters the resonance frequency
to $f=([k+k_{ts}]/m)^{1/2}/(2\pi)$. When the oscillation amplitude
of the cantilever is large, $k_{ts}$ can vary significantly within
one oscillation cycle, requiring averaging (see
text).\label{FigCLme}}
\end{figure}

\begin{figure}
\centering\includegraphics[width=14cm]{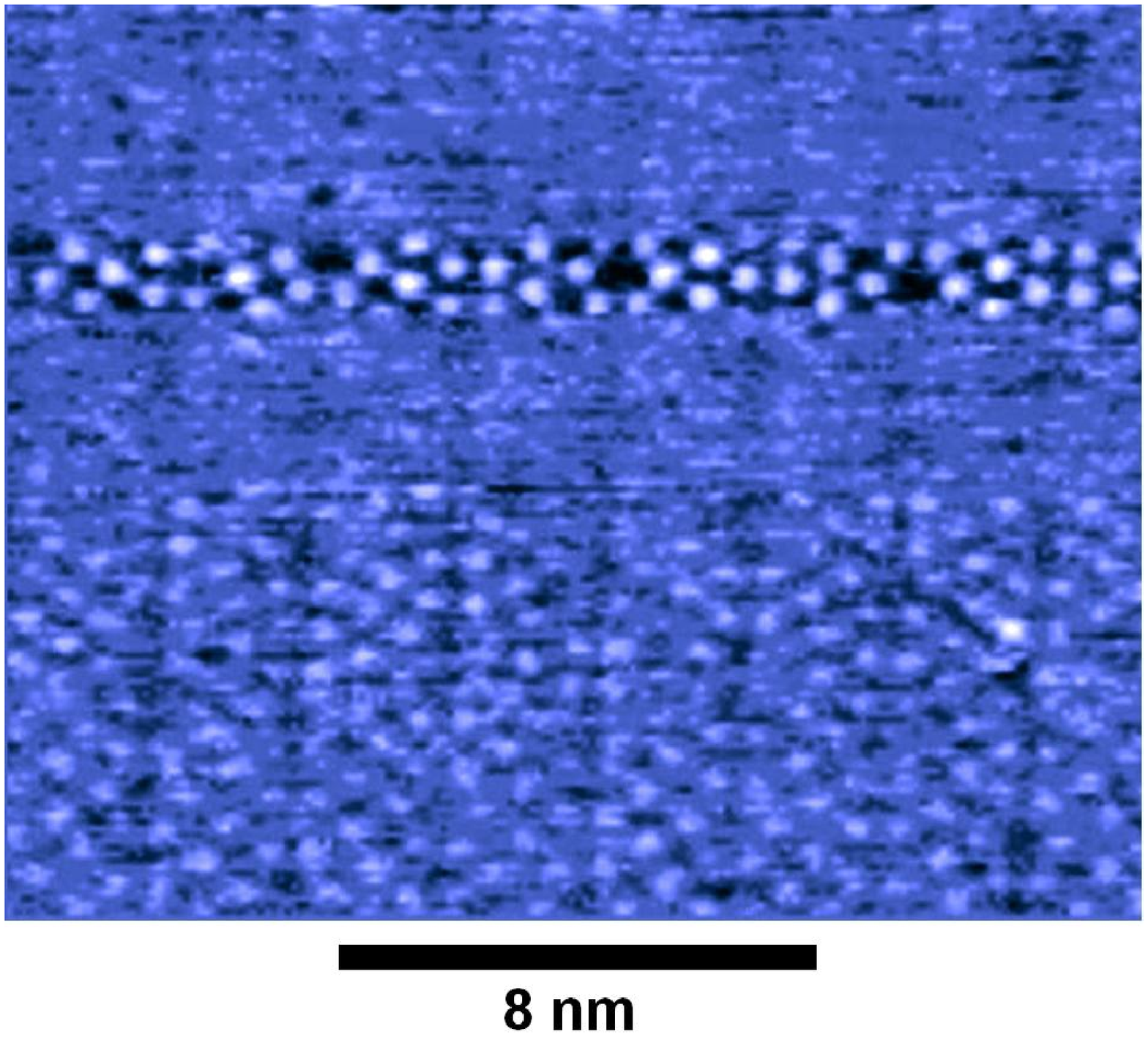} \caption{First
AFM image of a reactive surface showing true atomic resolution:
Si(111)-(7$\times$7) reconstruction. Parameters: $k=17$\,N/m,
$A=34$\,nm, $f_0=114$\,kHz, $\Delta f=-70$\,Hz and $Q=28\,000$,
scanning speed = 3.2\,lines/s. Environment: ultra-high vacuum,
room temperature. \cite{Giessibl:1995}\label{fig1st7x7}}
\end{figure}

\begin{figure}[h]
  \centering \includegraphics[width=16cm,clip=true]{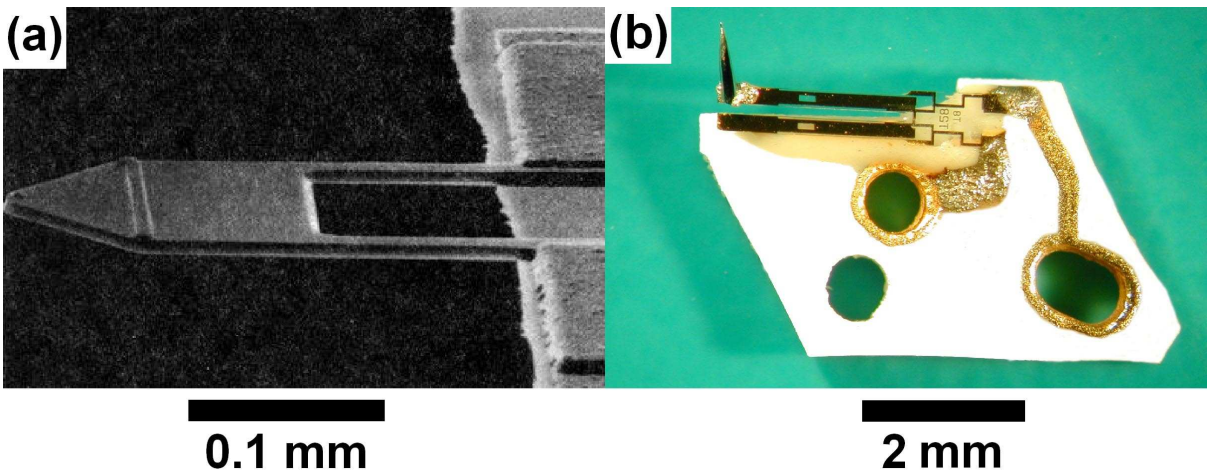}
  \caption{Micrographs of (a) a piezoresistive cantilever \cite{Tortonese:1993} and (b) a \lq qPlus\rq{} sensor\cite{Giessibl:2000b} - a cantilever made
  from a quartz tuning fork. The piezoresistive cantilever has a length of 250\,$\mu$m, a width of 50\,$\mu$m and a
  thickness of 4\,$\mu$m. The eigenfrequency is 114\,kHz, the stiffness 17\,N/m and the $Q$-factor in vacuum 28\,000.
  The qPlus sensor has a typical eigenfrequency ranging from 10 to 30\,kHz (depending on the mass of the tip), a
  stiffness of 1\,800\,N/m and a $Q$-factor of 4000 in vacuum at $T=300$\,K and 20\,000 at $T=4$\,K. One of the prongs is fixed to a large
  substrate and a tip is mounted to the free prong. Because the fixed
  prong is attached to a heavy mass, the device is mechanically
  equivalent to a traditional cantilever. The dimensions of the free
  prong are: Length: 2.4\,mm, width: 130\,$\mu$m, thickness:
  214\,$\mu$m.\label{figPLqP}}
\end{figure}

In amplitude modulation AFM \cite{Martin:1987}, the cantilever is
driven at a constant frequency and the vibration amplitude is a
measure of the tip-sample interaction. In 1991, Albrecht et al.
have shown that frequency modulation (FM)
AFM\cite{Albrecht:1991} offers even less noise at
larger bandwidth than amplitude modulation AFM. In FM-AFM, a
cantilever with a high quality factor $Q$ is driven to oscillate
at its eigenfrequency by positive feedback with an electronic
circuit that keeps the amplitude $A$ constant. A cantilever with a
stiffness of $k$ and effective mass $m$ has an eigenfrequency
given by $f_0=1/(2\pi)\sqrt{k/m}$. When the cantilever is exposed
to a tip-sample force gradient $k_{ts}$, its frequency changes
instantly to $f=f_0+\Delta f=1/(2\pi)\sqrt{k'/m}$ with
$k'=k+k_{ts}$ (see Fig. \ref{FigCLme}). When $k_{ts}$ is small
compared to $k$, the square root can be expanded and the frequency
shift is simply given by\cite{Albrecht:1991}
\begin{equation}\label{dfsA}
\Delta f(z)=\frac{f_{0}}{2k} k_{ts}(z).
\end{equation}
This formula is only correct if $k_{ts}$ is constant over the
distance range from $z-A$ to $z+A$ that is covered by the
oscillating cantilever. The force gradient $k_{ts}$ was probably
almost constant within the oscillation interval in the first
application of FM-AFM in magnetic force microscopy by Albrecht et
al.\cite{Albrecht:1991}, where magnetic recording media with
magnetic transitions spaced by about 2\,$\mu$m were imaged with a
cantilever with a stiffness of about 10\,N/m oscillating at an
amplitude of about 5\,nm. In contrast, in the newer application of
FM-AFM in atomic-resolution AFM, the force gradient varies by
orders of magnitude throughout the oscillation of the cantilever.
Using frequency modulation AFM, true atomic resolution on
Si(111)-(7$\times$7), a fairly reactive sample, could be achieved
in 1994 \cite{Giessibl:1995}. Figure \ref{fig1st7x7} shows the
topographic image of this data. The fast scanning direction was
horizontal, and the atomic contrast is rather poor in the lower
section, quite good in a narrow strip in the center part and
vanishing in the top section. These changes in contrast were due
to tip changes, indicating fairly strong interaction during the
imaging process. A piezoresistive cantilever made of Si\cite{Tortonese:1993}
as shown in Fig. \ref{figPLqP}(a) with a stiffness of $17$\,N/m
was used to obtain this image. The amplitude of the cantilever can
be freely adjusted by the operator, and while it was even planned
to use the thermally excited amplitude\cite{Giessibl:1994b}
($\approx 10$\,pm), the empirically determined optimal amplitude
values were always around 10\,nm -- on a similar order of
magnitude as the value of $A=34$\,nm used in the data of Fig.
\ref{fig1st7x7}. The chemical bonding forces that are responsible
for the atomic contrast in imaging Si by AFM have a range on the
order of 100\,pm,\cite{Perez:1997} and the amplitude was 340 times
as large. The requirement of such a large amplitude is in stark
contrast to intuition. Imagine an atom magnified to a size of an
orange with a diameter of 8\,cm. The range of the bonding force is
then only 4\,cm or so, and the front atom of the cantilever would
approach from a distance of 20\,m and only in the last few
centimeters of its oscillation cycle would feel the attractive
bonding forces from the sample atom next to it. On the other hand,
force gradients can be quite large in chemical bonds. According to
the well-known Stillinger-Weber potential\cite{Stillinger:1985}, a
classic model potential the interaction of Si atoms in the solid
and liquid phases, a single bond between two Si atoms has a force
gradient of $k_{ts}\approx +170$\,N/m at the equilibrium distance
of $z=235$\,pm and $k_{ts}\approx -120$\,N/m when the two Si atoms
are at a distance of $z=335$\,pm. Because of the relatively large
values of interatomic force gradients, even cantilevers with a
stiffness on the order of 1\,kN/m should be subject to significant
frequency shifts when oscillating at small amplitudes (page 5 in
Ref \cite{Giessibl:1996}).

Nevertheless, the large-amplitude FM technique has celebrated
great successes by imaging metals, semiconductors and insulators
with true atomic resolution
\cite{Morita:2002,Meyer:2003,Garcia:2002,Hofer:2003a,Giessibl:2003}.

\section{The search for optimal imaging parameters}

In order to understand why these large oscillation amplitudes were
necessary, a quantitative analysis of the physics of large
amplitude FM-AFM was necessary, starting with a calculation of
frequency shift for large amplitudes. If $k_{ts}$ is not constant
over one oscillation cycle, Eq. \ref{dfsA} no longer holds and
perturbation theory such as the Hamilton-Jacobi
theory\cite{Goldstein:1980} can be used to find the relationship
between frequency and tip sample forces.\cite{Giessibl:1997b}
Other perturbative approaches have confirmed the
result,\cite{Baratoff:1997,Duerig:1999a,Duerig:1999b,Livshits:1999}
and an instructive representation of the formula is
\begin{equation}\label{dflA}
\Delta f(z)=\frac{f_{0}}{\pi k} \int_{-1}^{1} k_{ts}(z-u
A)\sqrt{1-u^2}du.
\end{equation}

This equation is key to a physical understanding of FM-AFM
allowing to evaluate the impact of various force components on
$\Delta f$, the experimental observable. On a first glance, the
large-amplitude result resembles Eq. \ref{dfsA} where $k_{ts}(z)$
is replaced by an averaged value. The average force gradient is
computed by convoluting $k_{ts}(z)$ in the interval $z-A$ to $z+A$
with a semi-spherical weight function. The weight function has its
maximum at $u=0$ -- a distance $A$ away from the minimal
tip-sample distance. The minimal tip sample distance $z_{min}$ is
an important parameter in any STM or AFM experiment, because while
a small value of $z_{min}$ is desirable for optimal spatial
resolution, both tip and sample can be damaged if $z_{min}$ is too
small. We can now ask, if we keep $z_{min}$ constant and vary $A$,
what happens to our signal, the frequency shift $\Delta f$? The
answer is given in Eq. \ref{dflA}: as long as the gradient of the
tip-sample interaction $k_{ts}$ remains constant as the tip of the
cantilever moves over a $z$-range from $z_{min}$ to $z_{min}+2A$,
$\Delta f$ stays constant. However, as $A$ reaches the decay
length $\lambda$ of the interaction, the frequency shift drops
sharply at a rate $\propto (\lambda/A)^{3/2}$. It turns out that
for amplitudes larger than $\lambda$, $\Delta f$ is no longer
proportional to the force gradient, but to the product of force
and the square root of $\lambda$\cite{Giessibl:2000a} (or,
equivalently to the geometric average between potential and
force\cite{Ke:1999}). In FM-AFM with amplitudes large compared to
the interaction range, it is useful to define a quantity
$\gamma=\Delta f kA^{3/2}/f_{0}$\cite{Giessibl:1997b}. The
'normalized frequency shift' $\gamma$ connects the physical
observable $\Delta f$ and the underlying forces $F_{ts}$ with
range $\lambda$, where $\gamma\approx 0.4 F_{ts} \lambda^{1/2}$
(see Eqs. 35-41 in \cite{Giessibl:2003}). For covalent bonds, the
typical bonding strength is on the order of -1\,nN with $\lambda
\approx 1$\,\AA, resulting in $\gamma \approx
-4$\,fN$\sqrt{\textrm{m}}$, where a negative sign indicates
attractive interaction. The crossover from the small-amplitude
approximation in Eq. \ref{dfsA} to the large-amplitude case in Eq.
\ref{dflA} occurs for amplitudes on the order of the interaction
range $\lambda$.

Equation \ref{dflA} determines the influence of the oscillation
amplitude on AFM challenge number 3 (disturbing contribution of
long-range forces) outlined in the second section: Imagine an AFM
tip at a minimal distance $z_{min}=0.3$\,nm to a surface, where the total
tip-sample force is composed of a
chemical bonding force with an exponential distance dependence and a given
range with a long-range force with the same strength and a ten times
longer range (see caption of Table \ref{table1} for details). In large-amplitude AFM
(here, $A>1$\,nm), the signal is proportional to the normalized
frequency shift $\gamma$, and the long-range contribution to
$\Delta f$ is $\sqrt{1\,\textrm{nm}/100\,\textrm{pm}}\approx
3$-times larger than the short-range contribution. For small
amplitudes (here, $A<100$\,pm), $\Delta f$ is proportional to the
force gradient and the long-range component is only
$100\,\textrm{pm}/1\,\textrm{nm} = 1/10$ of the short-range
contribution. Therefore, small amplitude AFM helps to reduce the
unwanted contribution of long-range forces.

Even stronger attenuation of the unwanted long-range contribution
would be possible if higher order force derivatives could be mapped
directly. For example, if we could directly measure $\partial
^2F_{ts}/\partial z^2$, the long range component was only 1/100 of
the short-range contribution, and for a direct mapping of the
third order gradient $\partial ^3F_{ts}/\partial z^3$, the relative long
range component would reduce to a mere 1/1000. Higher force
gradients can be mapped directly by higher-harmonic AFM, as shown
further below.

\begin{table} [h]
\small
\begin{tabular}{|c||c|c|c|c|}
\hline AFM       & physical                              & short-range &long-range& relative short-    \\
  method      & observable                              & contribution &contribution&range contribution    \\
\hline\hline
quasistatic         & force                               & 1\,nN       & 1\,nN &50\,\%\\ \hline
large amplitude FM & $\gamma \approx 0.4\times$force$\times \sqrt{\textrm{range}}$ & 4\,fN$\sqrt{\textrm{m}}$     & 12\,fN$\sqrt{\textrm{m}}$ &25\,\%\\ \hline
small amplitude FM & force gradient                      & 10\,N/m     & 1\,N/m &91\,\% \\ \hline
higher-harmonic     & $n$-th force gradient               & $10^{n+9(n-1)}$\,N/m$^n$ & $10^{9(n-1)}$\,N/m$^n$ &$\approx 100$\,\%$(1-10^{-n})$ \\ \hline
\end{tabular}
\caption{Short- and long-range contributions to AFM signals in
different operating modes. This model calculation assumes a
chemical bonding force $F(z)=F_0 e^{-z/\lambda}$ with a strength
of $F_{short\,range}(z_{min})=1$\,nN and a range of
$\lambda_{short\,range}=100$\,pm and an equally strong long-range
background force with $F_{long\,range}(z_{min})=1$\,nN and a range
of $\lambda_{long\,range}=1$\,nm. Depending on the mode of AFM operation,
the short-range part has a different weight in the total interaction signal.
Higher-harmonic AFM offers the greatest attenuation of long-range forces.} \label{table1}
\end{table}

Because the forces that act in AFM are small, optimizing the
signal-to-noise ratio is crucial for obtaining good images.
Frequency noise in FM-AFM is inversely proportional to
amplitude\cite{Martin:1987,Albrecht:1991,Giessibl:2003,Hasegawa:2004}.
As discussed above, the signal stays constant until $A$ reaches
$\lambda$ and drops proportional to $(\lambda/A)^{3/2}$ for larger
amplitudes. Therefore, the signal-to-noise ratio is maximal
for amplitudes on the order of the decay length of the interaction
that is used for imaging.\cite{Giessibl:1999a} For atomic imaging,
amplitudes on the order of 100\,pm are expected to be optimal.

\begin{figure}[h]
  \centering \includegraphics[width=14cm,clip=true]{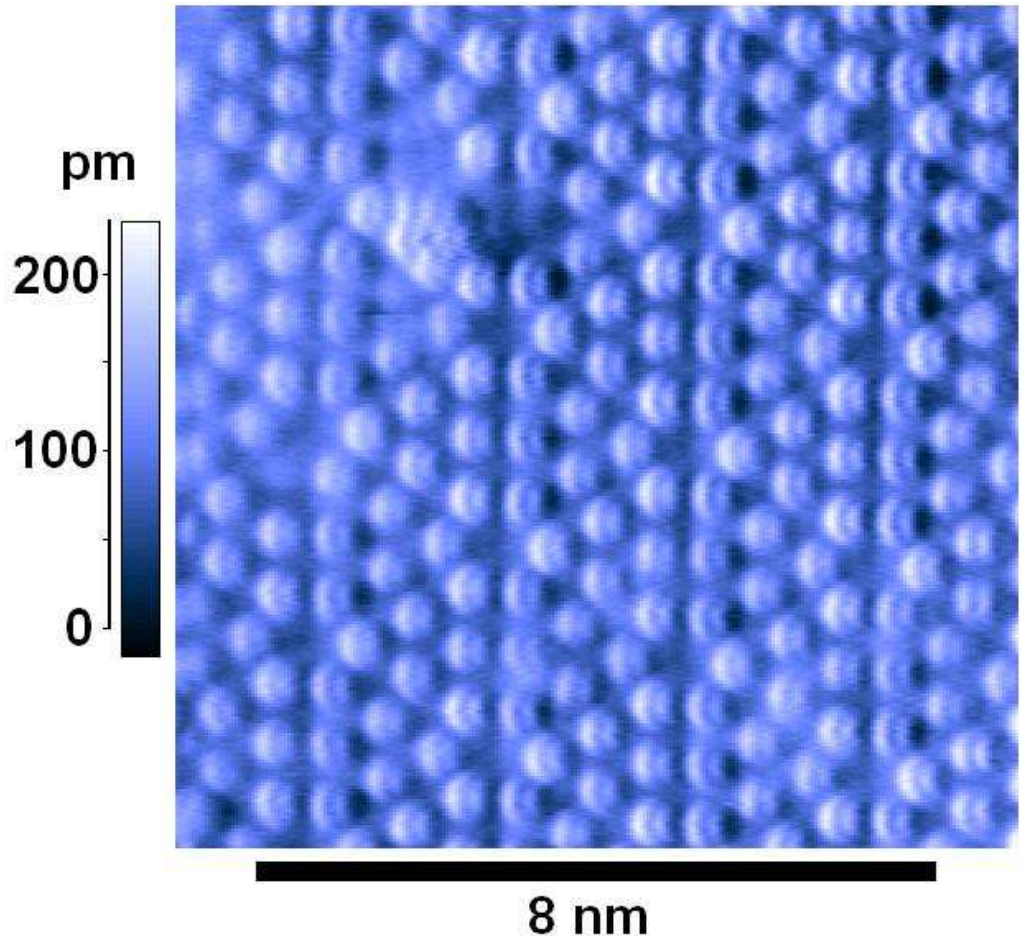}
  \caption{AFM image of the silicon 7$\times$7
reconstruction with true atomic resolution with a stiff
cantilever. Parameters: $k=1800$\,N/m, $A=0.8$\,nm,
$f_0=16.86$\,kHz, $\Delta f=-160$\,Hz and $Q=4\,000$. Environment:
ultra-high vacuum, room
temperature.\cite{Giessibl:2000c}\label{fig1stsubatomic}}
\end{figure}

\begin{figure}[h]
  \centering \includegraphics[width=14cm,clip=true]{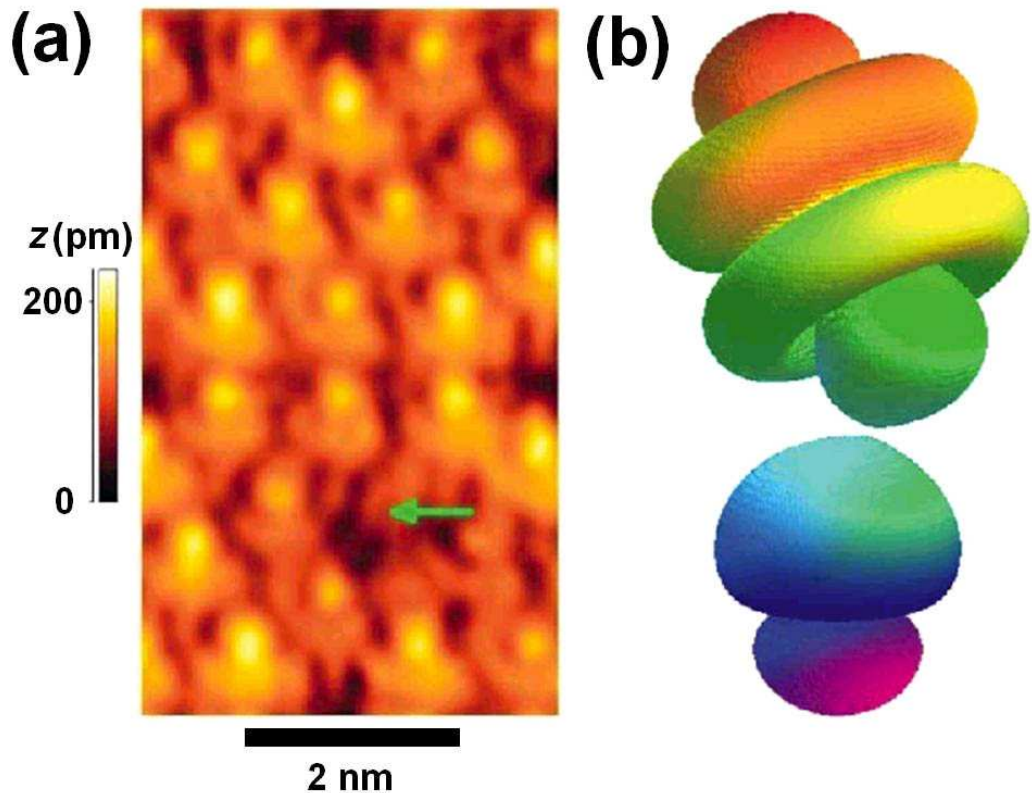}
  \caption{(a) Dynamic STM image of the silicon 7$\times$7
reconstruction where a Co$_6$Fe$_3$Sm tip was mounted on a qPlus
sensor. Parameters:
  $k=1800$\,N/m, $A=0.5$\,nm, $f_0=19\,621$\,Hz, sample bias voltage -100\,mV, average tunneling current 200\,pA.
  (b) Schematic plot of tip and sample states that can lead to the experimental image shown in (a). The sample state is a
  dangling bond of a Si adatom with its 3sp$^3$ symmetry, while a Sm 4f$_{z^3}$ state is taken as a tip state. Environment: ultra-high
vacuum, room temperature.\cite{Herz:2003b}\label{figsamarium}}
\end{figure}

As a conclusion of these calculations, we find that the use of
small amplitudes $A\approx \lambda$ would have two advantages:
\begin{enumerate}
    \item Increased signal-to-noise ratio\cite{Giessibl:1999a}
    \item Greater sensitivity to short-range
    forces\cite{Giessibl:2003}.
\end{enumerate}
So why was it not feasible to use small amplitudes in the initial
experiments? Two reasons, related to the mechanical stability of
the oscillating cantilever, can be identified. First,
jump-to-contact is prevented if the withdrawing force of the
cantilever when it is closest to the sample given by $k\times A$
is larger than the maximal attraction.\cite{Giessibl:1997b}
Second, because tip-sample forces are not
conservative\cite{Kantorovich:2001b}, random dissipative phenomena
with a magnitude of $\delta E_{ts}$ cause amplitude fluctuations
$\delta A = \delta E_{ts}/(kA)$
\cite{Giessibl:1999a,Giessibl:2004}. Both problems can be resolved
by utilizing cantilevers with sufficient stiffness. Stability
considerations propose a lower threshold for $k$ that depends on
the tip-sample dissipation as well as the $Q$-factor of the
cantilever. Because the frequency shift is inversely proportional
to the stiffness (Eqs. \ref{dfsA} and \ref{dflA}), $k$ should
still be chosen as low as permitted by the stability requirements.
Stiff cantilevers were not commercially available when we realized
their potential advantages, therefore we built cantilevers with a
stiffness of $k=1800$\,N/m from quartz tuning
forks\cite{Giessibl:1998,Giessibl:2000b} (see Fig.
\ref{figPLqP}(b)). A secondary advantage of quartz cantilevers is
their greater frequency stability with temperature, which leads to
lower frequency drift in particular if a quartz stabilized frequency detector
is used (we used the EasyPLL by Nanosurf AG, Liestal, Switzerland).
Other small-amplitude approaches with stiff
home-built tungsten cantilevers have been demonstrated in Ragnar
Erlandsson's\cite{Erlandsson:1996} and John Pethica's
groups.\cite{Hoffmann:2001,Hoffmann:2001a,Oral:2001a} As predicted
by theoretical considerations, the stiff cantilever allowed to use
sub-nm amplitudes, resulting in an improved signal-to-noise ratio,
a strong attenuation of the disturbing long-range forces and the
possibility of stable scanning at very small tip-sample distances.
For these reasons, the spatial resolution was increased as shown
in Fig. \ref{fig1stsubatomic}. The image shows a very clear
picture of Si with a defect and very large corrugation. The
adatoms of Si which should be spherically symmetric showed
subatomic details that are interpreted as orbitals in the tip
atom\cite{Giessibl:2000c,Huang:2003}. This AFM image seemed to
show greater resolution than what was known from STM. According to
the \lq Stoll-formula\rq{} \cite{Stoll:1984}, a theoretical
estimate of the vertical corrugation and thus the lateral resolution
of STM images, two physical parameters
are crucial for the high spatial resolution of STM: a) the very short decay length
of the tunneling current and b) a small tip-sample distance. Three
likely reasons have been identified that may explain why dynamic
AFM could provide better resolution than STM:\cite{Giessibl:2001d}
\begin{enumerate}
    \item In dynamic AFM, the minimal tip sample distance can be
    much smaller than in STM without destroying the tip, because the shear forces that act on
    the front atom during scanning are much smaller in the oscillation phase where the tip is far from the sample.
    \item When using large gap voltages, a variety of states can
    contribute to the tunneling current, smearing out the image.
    \item Tip-sample forces also have repulsive components with a
    very short decay length.
\end{enumerate}
The first two characteristics can be fulfilled in STM as well by
using a very small tunneling bias voltage and oscillating the STM
tip similar to an AFM tip. Figure \ref{figsamarium} shows an image
obtained in dynamic STM where a Co$_6$Fe$_3$Sm magnetic tip was
mounted onto a qPlus sensor, imaging
Si.\cite{Herz:2003a,Herz:2003b} Each Si adatom looks like a fried
egg with a sharp center peak surrounded by a halo. The radius of
the center peak is only on the order of 100\,pm, showing that
higher-momentum states\cite{Chen:1993} must have been involved in
this image. The experiment was repeated with pure Co, Fe and Sm
tips, and only pure Sm tips yielded similar images as Figure
\ref{figsamarium}, we therefore concluded that a Sm atom acted as
the tip atom in this experiment.\cite{Herz:2003a} In atomic
samarium, the electrons at highest occupied state are in a 4f
state. If one assumes, that the electronic states at a Sm surface
atom of bulk Co$_6$Fe$_3$Sm are similar to atomic states in Sm, it
appears likely that the crystal field around the front atom
creates a state close to 4f$_{z^3}$ symmetry that is responsible
for the tunneling contrast. Interestingly, very small tip-sample
distances could only be realized with oscillating tips. When the
oscillation was turned off, the current setpoint had to be reduced
otherwise the tip would not survive the small tunneling distances.

\begin{figure}[h]
  \centering \includegraphics[width=14cm,clip=true]{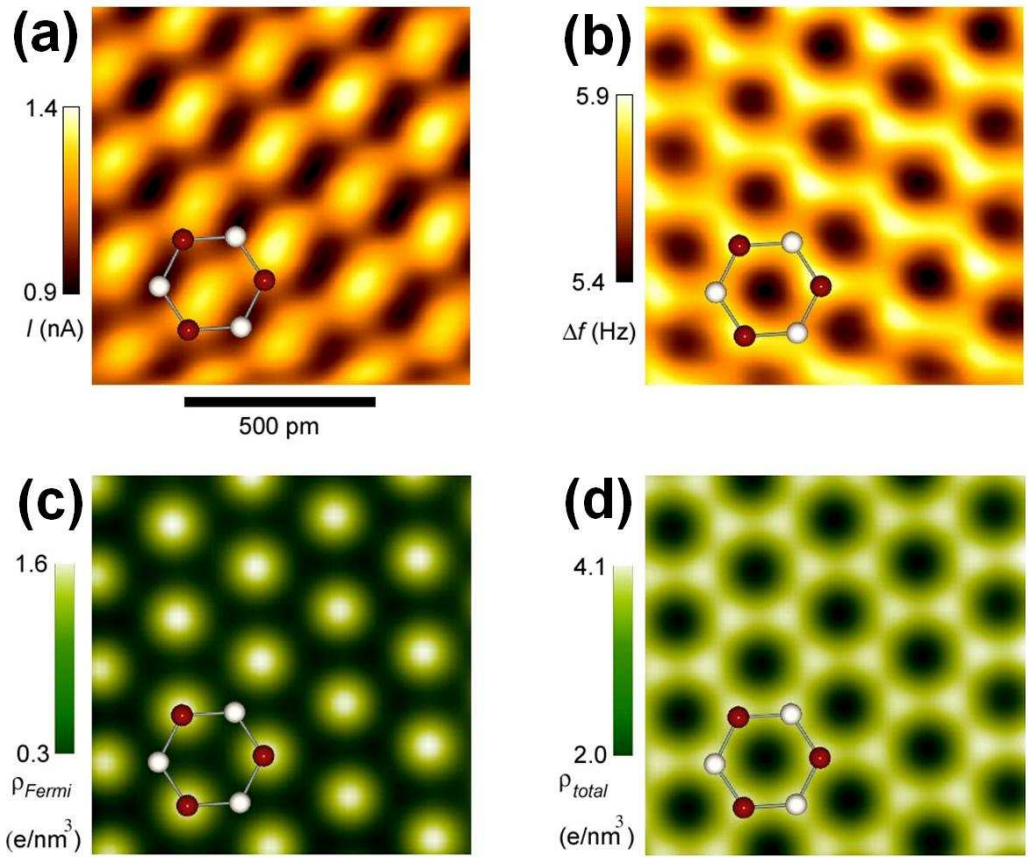}
  \caption{(a) Constant height STM image of graphite and (b) a simultaneously recorded AFM
  image (repulsive). Part (c) shows an estimate of the charge density at the Fermi level (visible by STM) and
  (d) the total charge density (relevant for repulsive AFM) for graphite. Parameters:
  $k=1800$\,N/m, $A=0.3$\,nm, $f_0=18076.5$\,Hz, and $Q=20\,000$. \cite{Hembacher:2003}\label{figsimIF}}
\end{figure}

Operation at small oscillation amplitudes not only results in
greater resolution, it also facilitates simultaneous STM and AFM
imaging. A straightforward implementation of combined current- and
force measurements uses the constant-height mode, where the
$z$-position of the tip is held constant relative to the plane
connecting the surface atoms. A simultaneous measurement of
tunneling current and frequency shift allows to compare forces and
tunneling currents. Figure \ref{figsimIF} shows a comparison of
current and repulsive force on graphite\cite{Hembacher:2003}
observed by simultaneous AFM and STM in vacuum at liquid helium
temperatures (4.9\,K). STM only sees the electrons at the Fermi
level, while repulsive forces act wherever the local charge
density is high (i.e. over \emph{every} atom) for small enough distances. In graphite, only
every second surface atom conducts electricity, but every surface
atom exerts repulsive forces. Therefore, AFM \lq\lq sees more\rq\rq{} than STM
and allows to correlate topography to local conductance. This
method is promising for other materials with more than one basis
atom in the elementary cell. The images have been taken with a
low-temperature AFM/STM operating at 4.9\,K in ultra-high
vacuum.\cite{Hembacher:2003b,Hembacher:2005}

\begin{figure}[h]
  \centering \includegraphics[width=14cm,clip=true]{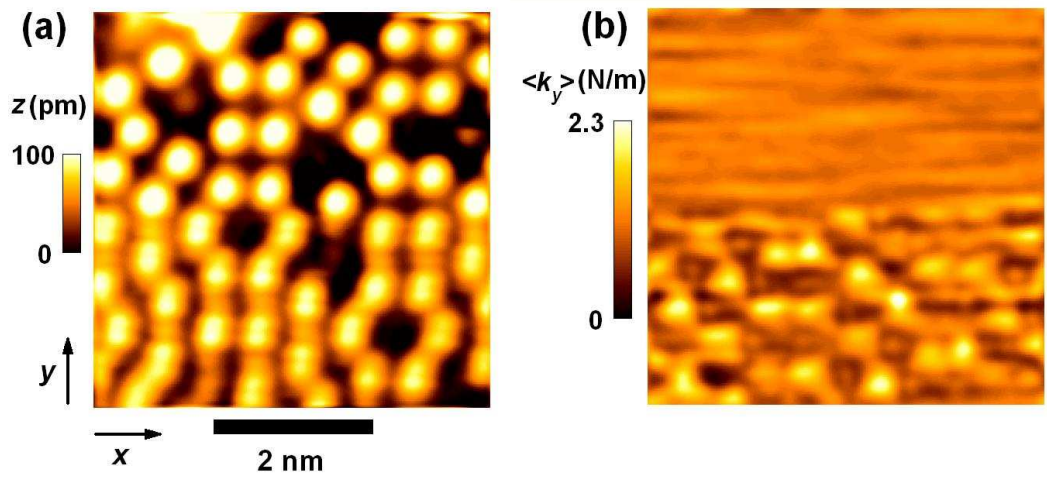}
  \caption{(a) Topographic STM image of Si(111)-(7$\times$7)
where the tip is mounted on a lateral force sensor. The tip
oscillates with $A\approx 80$\,pm in the $y$-direction in the
lower half of the image, the oscillation is turned off in the
upper half. (b) Corresponding lateral force gradient. On top of
the adatoms, the bond between tip and sample causes an increase in
frequency shift. Parameters:
  $k=1350$\,N/m, $A=80$\,pm (bottom), $A=0$ (top), $f_0=10214$\,Hz. Environment: ultra-high
vacuum, room temperature. \cite{Herz:2003b}\label{figLFM}}
\end{figure}

While a strong bias dependence holds both for atomic-resolution
STM\cite{Feenstra:1987} as well as AFM
images\cite{Arai:2004,Hembacher:2005}, one pronounced difference
is that the direction of the tunneling current is not accessible in STM,
while the direction of the measured force is determined by the orientation
of the cantilever. Usually, AFM senses forces that are normal to the
surface, but it is also possible to perform lateral force
microscopy\cite{Mate:1987} by measuring the forces acting parallel
to the surface. In a quasistatic mode, lateral forces can be
recorded simultaneously with normal forces. In dynamic modes, it
is easier to rotate the attachment of the cantilever by 90 degrees
and detect lateral forces. Figure \ref{figLFM} shows a measurement
of the lateral force gradients between a tip and a Si surface.
Parallel motion between tip and cantilever also allows to use
extremely soft cantilevers without suffering jump-to-contact to
probe the limits of force resolution, as demonstrated by Rugar et
al. in single spin detection by magnetic resonance force
microscopy \cite{Rugar:2004}.

\section{Higher-harmonic atomic force microscopy}
Can we increase the spatial resolution of AFM any further? When
decreasing the amplitude from $A>>\lambda$ to $A<<\lambda$, the
frequency shift changes from a proportionality of
$F_{ts}\sqrt{\lambda}$ to $F_{ts}/\lambda$. As outlined above, an
experimental observable that is proportional to a higher force
gradient should allow even higher spatial resolution than
small-amplitude FM-AFM. Luckily, there is a physical observable
that couples directly to higher force gradients. When the
cantilever oscillates in the force field of the sample, a shift in
frequency is not the only change in the cantilever's motions. The
oscillation of the cantilever changes from a purely sinusoidal
motion given by $q'=A\cos(2\pi f t)$ to an oscillation that contains higher
harmonics with $q'=\sum_{n=0}^\infty
a_n\cos(2\pi n f t+\phi_n)$. For amplitudes that are large with
respect to the range of $F_{ts}$, the higher harmonics are
essentially proportional to $\Delta f$ \cite{Duerig:1999b}.
However, for small amplitudes, D\ürig has found that $F_{ts}$
could be recovered immediately within the distance range from
$z_{min}$ to $z_{min}+2A$ if the amplitudes and phases of all
higher harmonics of the cantilever's motion were
known.\cite{Duerig:2000} Moreover, higher harmonics bear even more
useful information: direct coupling to higher force
gradients.\cite{Hembacher:2004} Similar to Eq. \ref{dflA}, we can
express the magnitude of the higher harmonics by a weighted
average of a force gradient -- a gradient of order $n>1$ this
time:
\begin{equation}\label{a_n}
    a_{n} =\frac{2}{\pi k} \frac{1}{1-n^2}
\frac{A^n}{1\cdot3\cdot...\cdot(2n-1)}
\int_{-1}^{1}\frac{d^{n}F_{ts}(z+A u)}{dz^{n}} (1-u^2)^{n-1/2} du.
\end{equation}
The weight function changes from the semi-spherical shape
$w_{\Delta f}(u)=(1-u^2)^{1/2}$ in Eq. \ref{dflA} to functions
$w_n(u)=(1-u^2)^{n-1/2}$ that are more and more peaked with
increasing $n$. For this reason, the use of small amplitudes is of
even greater importance in higher-harmonic AFM than in FM-AFM. The
magnitude of the higher harmonic amplitudes $a_n$ is rather small
compared to the fundamental amplitude $a_1=A$, therefore higher
harmonic AFM works best at low temperatures, where the detection
bandwidth can be set to very small values.

The spatial resolution of
AFM and STM is fundamentally neither limited by the mechanical vibration level
nor by thermal vibrations, but by the
spatial extent of the experimental objects that are observed --
electrons at the Fermi level in STM\cite{Tersoff:1985}, and
something close to the total charge density in repulsive
AFM\cite{Ciraci:1990}. When probing the resolution limits of AFM, we first have to find an
object with the desired sharply localized electronic states. Pauling\cite{Pauling:1957}
has noted, that transition metals show a covalent bonding
character, and should therefore expose lobes of increased charge
density towards their neighbors. Indeed, while the surface atoms
of W(001) expose a large blurred charge cloud at the Fermi level
for $k$-vectors perpendicular to the surface (Fig. 8 in
Ref.\cite{Posternak:1980}), the total charge density shows four
distinct maxima (Fig. 3 in Ref.\cite{Posternak:1980} and Fig. 3(a)
in Ref.\cite{Mattheiss:1984}). Figure \ref{fighh} shows a direct
comparison of the simultaneously recorded tunneling current and
the amplitudes of the higher harmonics. As expected, the higher
harmonic data shows much greater detail.

\section{Summary and Conclusion}
\begin{figure}[h]
  \centering \includegraphics[width=16cm,clip=true]{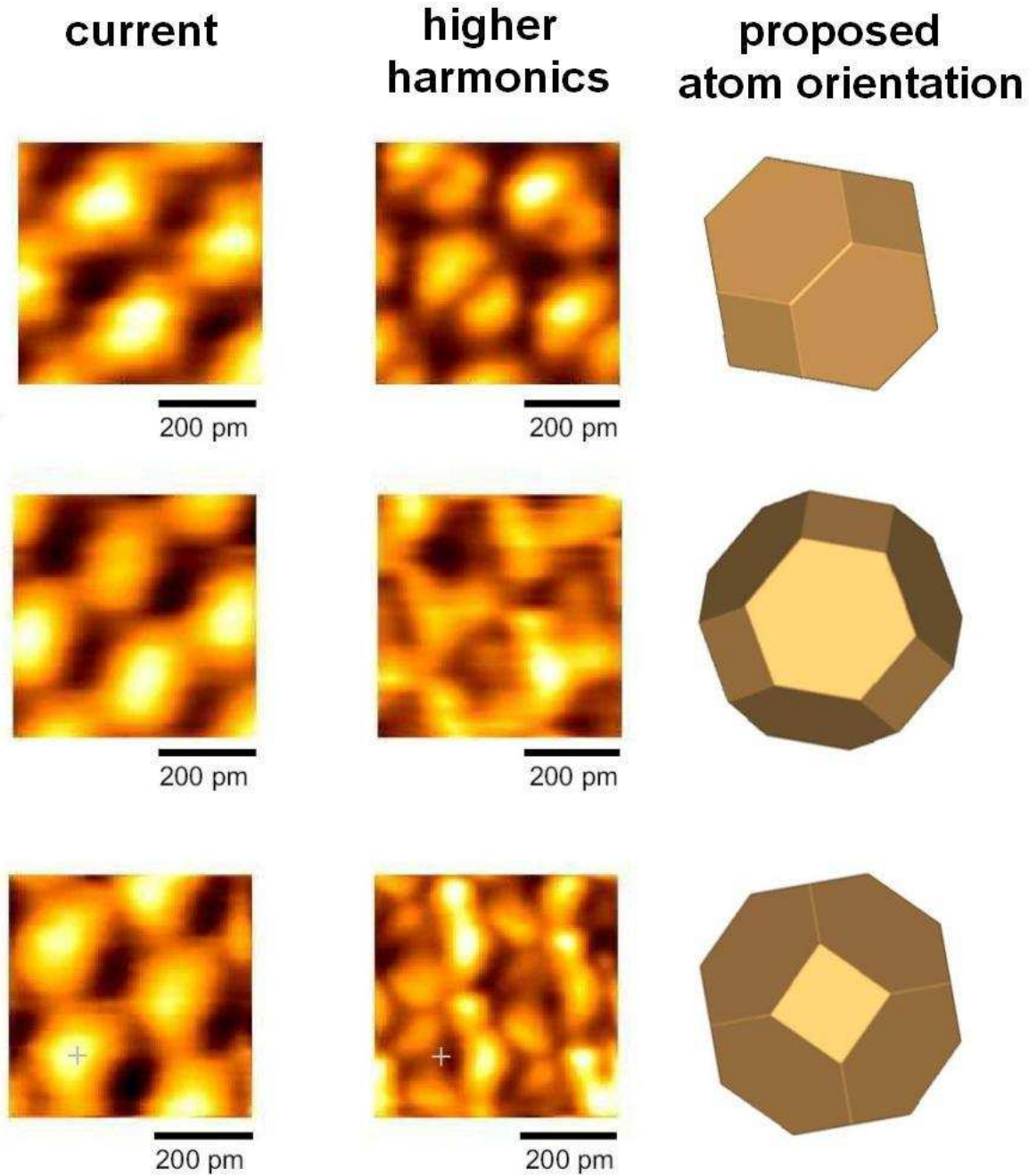}
  \caption{Simultaneous constant height STM (left column) and higher-harmonic AFM images
  (center column) of graphite with a tungsten tip.
  The right column shows the proposed orientation of the W tip
  atom. The W atom is represented by its Wigner-Seitz unit cell,
  which reflects the full symmetry of the bulk. We assume, that
  the bonding symmetry of the adatom is similar to the bonding
  symmetry of the bulk. This assumption is based on charge
  density calculations of surface atoms \cite{Posternak:1980,Mattheiss:1984}
  In the first row,
  the higher harmonics show a two-fold symmetry, as resulting from a [110] orientation of the
  front atom. In the second row, the higher harmonics show roughly a three-fold symmetry, as
  expected for a [111] orientation. In the third row, the symmetry of the higher-harmonic signal
  is approximately four-fold, as expected for a tip in [001] orientation. Parameters:
  $k=1800$\,N/m, $A=0.3$\,nm, $f_0=18\,076.5$\,Hz, and $Q=20\,000$. Environment: ultra-high
vacuum, $T=4.9$\,K.  \cite{Hembacher:2004}\label{fighh}}
\end{figure}

\begin{figure}[h]
  \centering \includegraphics[width=14cm,clip=true]{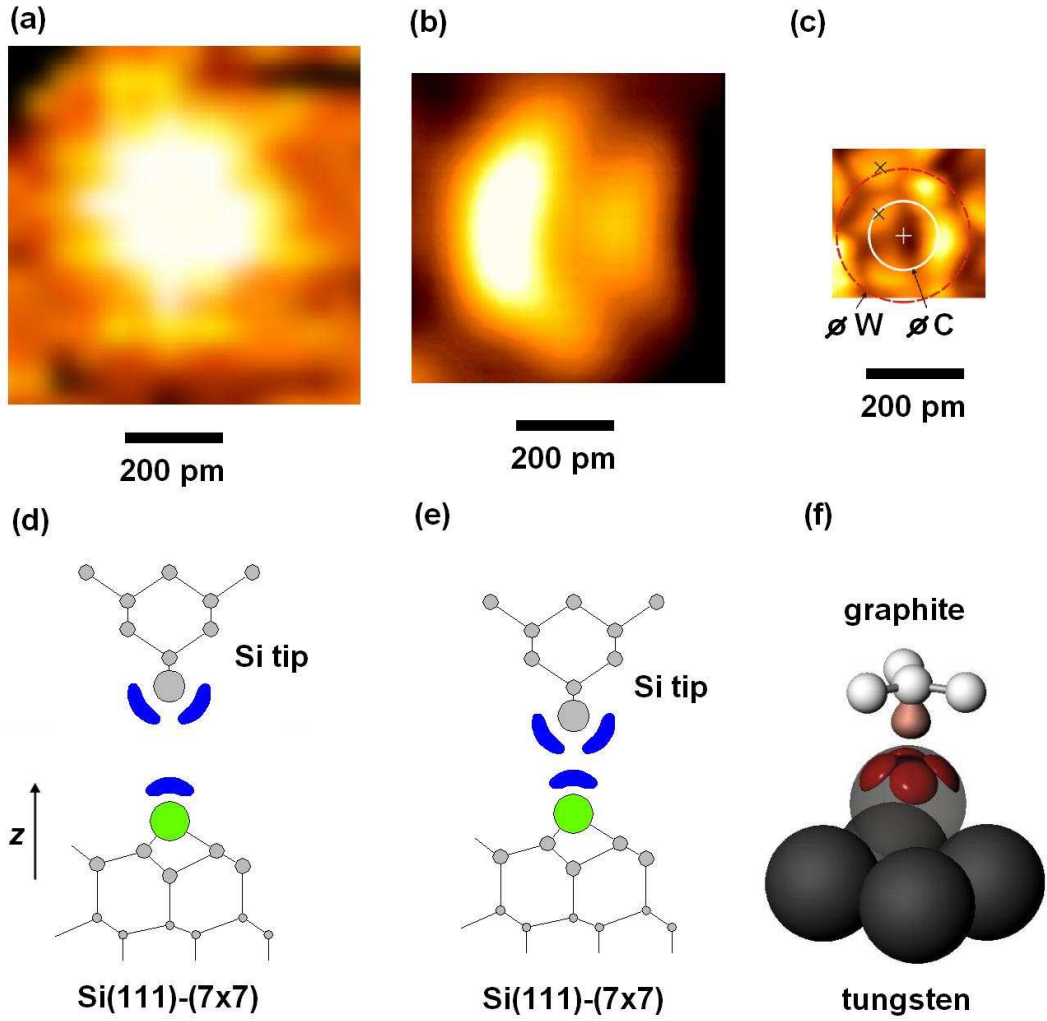}
  \caption{Progress in spatial resolution of AFM showing images of single
  atoms. The lateral scale in (a)-(c) is equal.
  (a) An adatom of the Si(111)-(7$\times$7) reconstruction, showing up as a blurred spot.
  (b) An adatom of the Si(111)-(7$\times$7) reconstruction, showing subatomic contrast originating in the electronic structure of the tip.
  (c) Higher-harmonic image of a tungsten atom mapped by a carbon atom.
  Parameters:
  (a) $k=17$\,N/m, $A=34$\,nm, $f_0=114$\,kHz, $\Delta f=-70$\,Hz and
  $Q=28\,000$ (ultra-high vacuum, room temperature),
  (b) $k=1800$\,N/m, $A=0.8$\,nm, $f_0=16\,860$\,Hz, $\Delta f=-160$\,Hz and $Q=4\,000$ (ultra-high vacuum, room temperature),
  (c) $k=1800$\,N/m, $A=0.3$\,nm, $f_0=18\,076.5$\,Hz, and $Q=20\,000$ (ultra-high vacuum, $T=4.9$\,K), higher harmonic detection.
  (d) Schematic view of a Si(001) tip close to a
  Si(111)-(7$\times$7) surface. Because of the large amplitude and
  a fairly large minimum tip-sample distance, the blurry image (a)
  corresponding to this configuration is approximately symmetric
  with respect to the vertical axis.
  (e) Similar to (d), but at a closer distance. The angular
  dependence of the bonding forces is noticeable.
  (f) W(001) surface close to a C atom in a graphite surface. The
  charge distribution in W shows small pockets that are resolved by higher-harmonic AFM
  with a light-atom carbon-probe.\label{figresolhist}}
\end{figure}

We have emphasized the enormous usefulness of AFM by referring to the numerous
references to the original publication\cite{Binnig:1986b} in the
introduction. While most AFM applications are currently not in the
atomic resolution regime, the enhancement in spatial resolution
is likely to add significant value in most AFM studies in physics,
chemistry, biology and materials science. Recently, true atomic
resolution by FM-AFM has been observed at ambient pressure in an
N$_2$ atmosphere \cite{Sasahara:2004}, showing that some of the
concepts of vacuum AFM are applicable in ambient environments.
Although
STM resolution can benefit from oscillating the tip, a concept that
has originated in AFM, Fig. \ref{fighh} shows that AFM has now
clearly reached and even surpassed the
resolution capability of STM. Figure \ref{figresolhist} shows the evolution of the resolution of AFM
from large amplitude AFM in 1994 (a) to small amplitude AFM in
2000 (b) and higher harmonic AFM in 2004 (c). While the structures
within single atoms shown in Fig. \ref{figresolhist} (b) and (c)
originate in the front atom of the probe,
other examples where AFM shows more atomic details of specimens than STM
such as the observation of the rest atoms in
Si(111)-(7$\times7$)\cite{Lantz:2000,Eguchi:2002} or the
observation of all dangling bonds on the Si/Ge(105)
surface\cite{Eguchi:2004} establish the improved spatial
resolution of AFM over STM in special cases. Atomic- and molecular structuring has
been the domain of STM for a long time, starting with the first
demonstration of manipulating single atoms\cite{Eigler:1990} to a
variety of nano-structuring methods by STM\cite{Rosei:2004}.
Recently, it has been shown that atomic manipulation by AFM is
possible even at room temperature\cite{Sugimoto:2005}.

We have not been able to discuss the phenomenal success of AFM in
biology, a field with a much more immediate impact on the human
condition. It can be expected that at least some of the concepts
that have been developed for AFM in vacuum will enable greater
resolution in biological AFM applications as well
\cite{Miles:1997,Horber:2003}.

{\bf Acknowledgments}

I wish to thank Jochen Mannhart for support and editorial
suggestions and my current and former students Martin Breitschaft,
Philipp Feldpausch, Stefan Hembacher, Markus Herz, Christian
Schiller, Ulrich Mair, Thomas Ottenthal and Martina Schmid for
their contributions towards the progress of AFM. I also thank Gerd
Binnig, Calvin F. Quate and Christoph Gerber for kicking off the fun of
the AFM field and for ongoing inspiring interactions. Special thanks to Heinrich Rohrer, Calvin Quate and
Christoph Gerber for critical comments and to German Hammerl for
help with LaTeX. Supported by the Bundesministerium f\ür Forschung
und Technologie under contract 13N6918.

\newpage

\bibliography{2005fjg}

\newpage
{\bf \Large Figures}

\end{document}